\documentclass{jpsj-suppl}
\usepackage{amssymb}
\usepackage{times}
\usepackage{color}
\usepackage{wrapfloat}
\def\cH{{\mathcal H}}
\def\eff{{\rm eff}}
\def\ave#1{\langle#1\rangle}  
\title{Edge Modes in the Intermediate-$D$ and Large-$D$ Phases of the $S=2$ Quantum Spin Chain
with $XXZ$ and On-Site Anisotropies}

\author{Kiyomi~Okamoto${}^1$\thanks{E-mail address: kokamoto@phys.titech.ac.jp}, 
Takashi~Tonegawa${}^{2,3}$,
T\^oru~Sakai${}^{4,5}$, 
and Makoto Kaburagi${}^2$ 
}
\inst{${}^1$Department of Physics, Tokyo Institute of Technology, Tokyo 152-8551, Japan \\
${}^2$Professor Emeritus, Kobe University, Kobe 657-8501, Japan \\
${}^3$Department of Physical Science,
School of Science, Osaka Prefecture University, Sakai 599-8531, Japan\\
${}^4$Japan Atomic Energy Agency, SPring-8, Hyogo 679-5148, Japan\\
${}^5$Graduate School of Material Science, University of Hyogo, Hyogo 678-1297, Japan
}
\abst{We investigate the edge modes at $T=0$ in the intermediate-$D$ (ID) phase and
the large-$D$ (LD) phase of the $S=2$ quantum spin chain with the XXZ anisotropy
and the generalized on-site anisotropies by use of the DMRG. 
There exists a gapless edge mode in the ID phase, while no gapless edge mode in the LD phase.
These results are consistent with the physical pictures of these phases.
We also show the ground-state phase diagrams obtained
by use of the exact diagonalization and the level spectroscopy analysis.}

\kword{quantum spin chain models, quantum phase transitions, critical phenomena, DMRG, edge modes}

\begin{document}
\maketitle

\section{Introduction}

In these years quantum spin chains have been attracting a great deal of attention
because rich physics is involved in spite of their apparent simplicity.
The most striking example is the existence of the Haldane phase \cite{haldane1,haldane2}
in the $S=1$ quantum spin chain only with nearest-neighbor (nn) isotropic interaction.

Very recently we \cite{tone1,oka1,oka2} have investigated the
$S=2$ quantum spin chain with the $XXZ$ and on-site anisotropies
described by
\begin{equation}
    \cH
    = \sum_j (S_j^x S_{j+1}^x + S_j^y S_{j+1}^y +  \Delta S_j^z S_{j+1}^z)
      + D_2 \sum_j (S_j^z)^2,
    \label{eq:ham1}
\end{equation}
where $S_j^\mu$ ($\mu = x,y,z$)  denotes the $\mu$-component of the $S=2$ spin operator
at the $j$-th site.
The quantity $\Delta$ represents the $XXZ$ anisotropy parameter of the nn interaction,
and $D_2$ the on-site anisotropy parameter.
The history of the works on the above model as of 2011 is summarized in \cite{tone1}.

\begin{figure}[ht]
   \begin{center}
       \scalebox{0.29}{\includegraphics{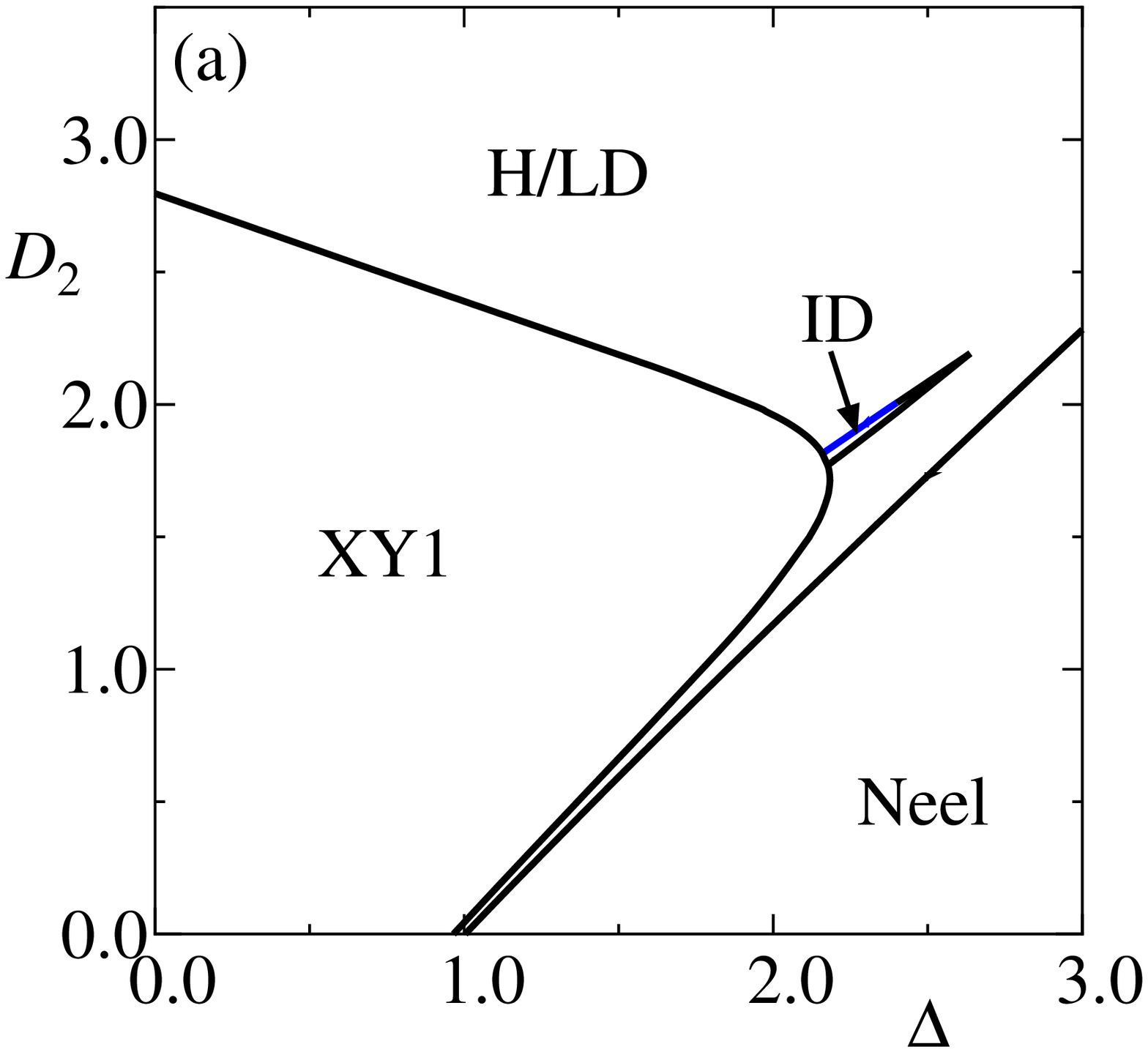}}~~~~
       \scalebox{0.29}{\includegraphics{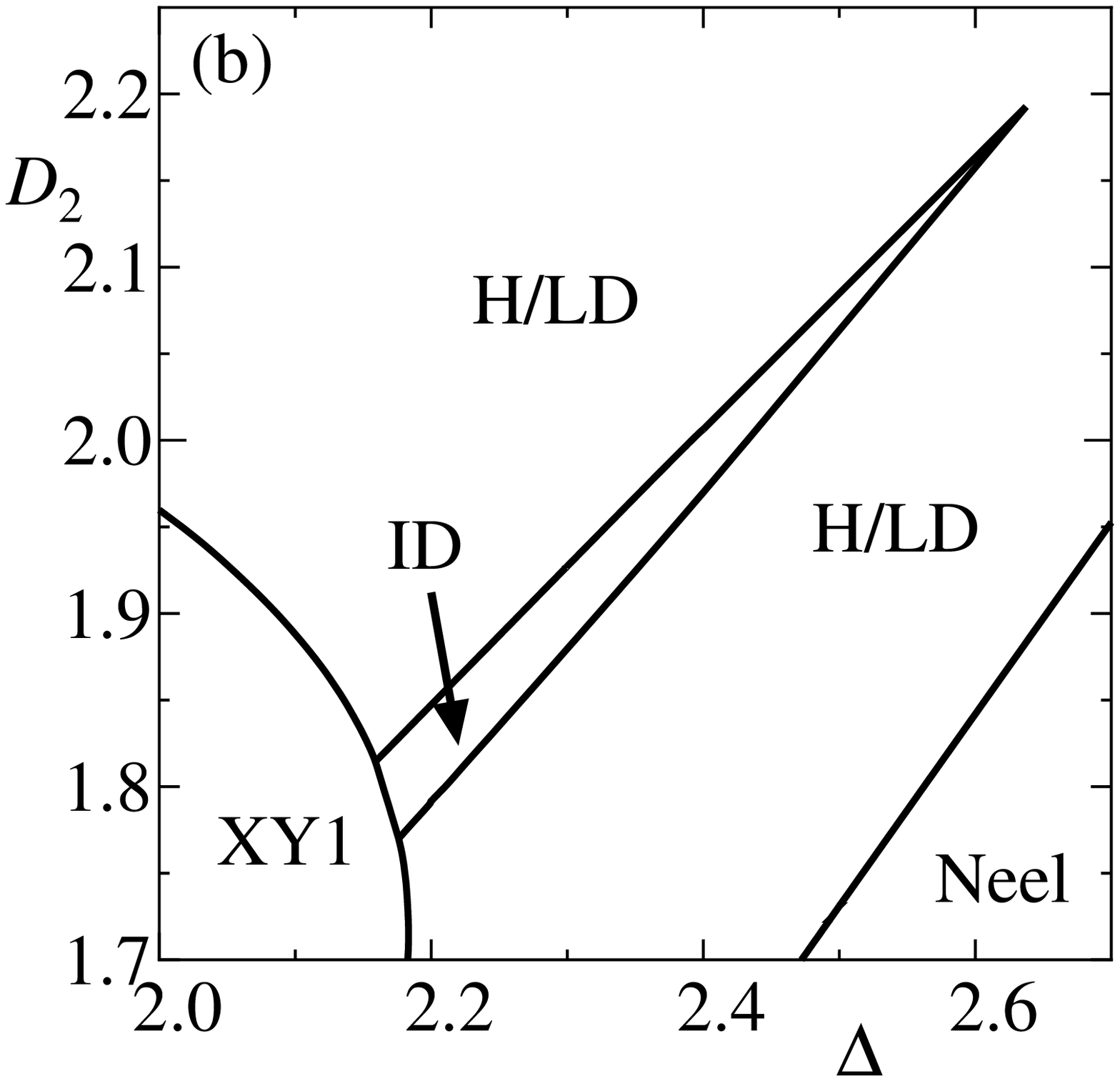}}
   \end{center}
   \caption{Phase diagram of Hamiltonian (\ref{eq:ham1}) \cite{tone1,oka1,oka2}.
            We have restricted ourselves to the case of  $\Delta \ge 0$ and $D_2 \ge 0$.
           } 
   \label{fig:pd1}
\end{figure}
\begin{figure}[h]
      \begin{center}
         \scalebox{0.25}[0.25]{\includegraphics{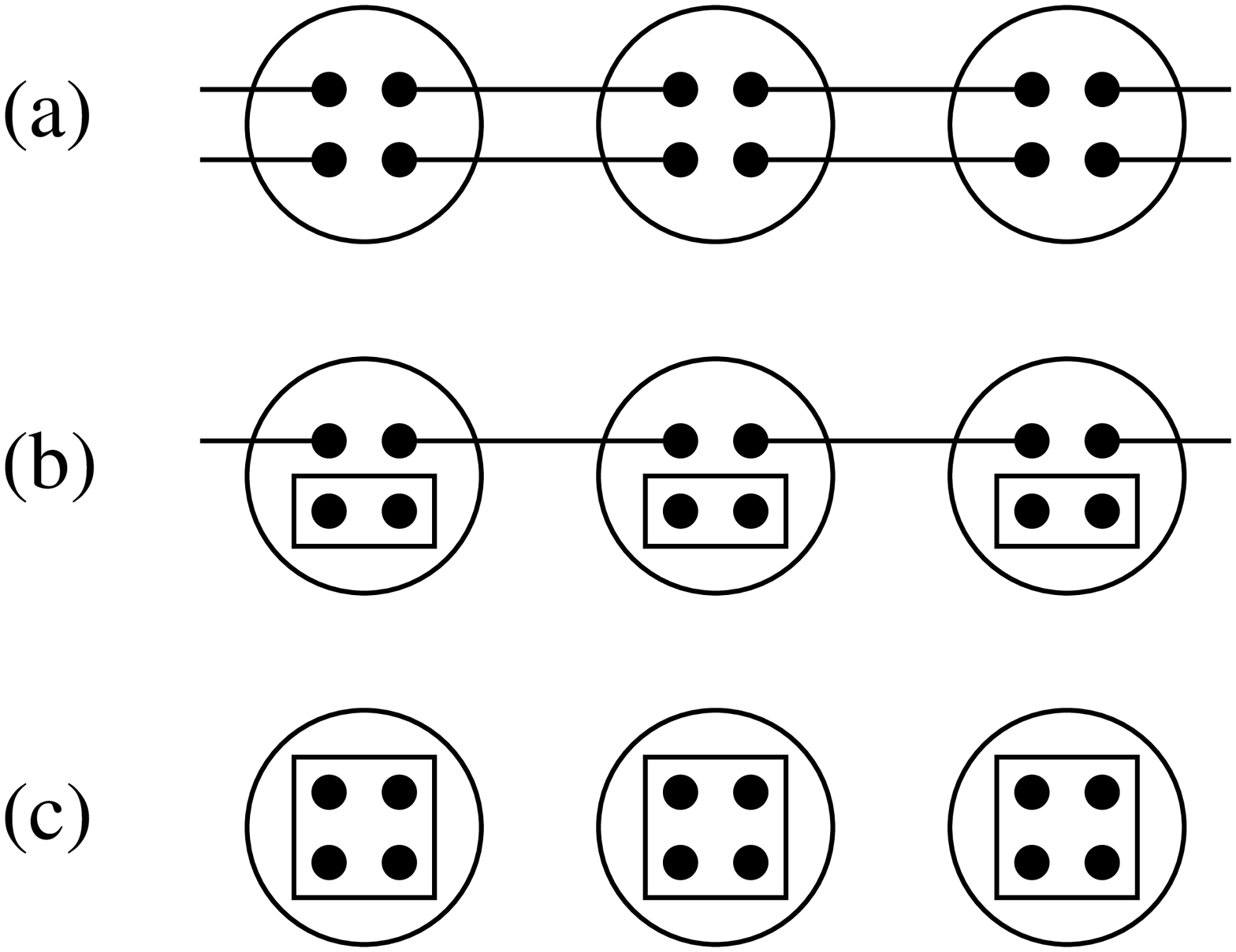}}~~~~~~
         \scalebox{0.25}[0.25]{\includegraphics{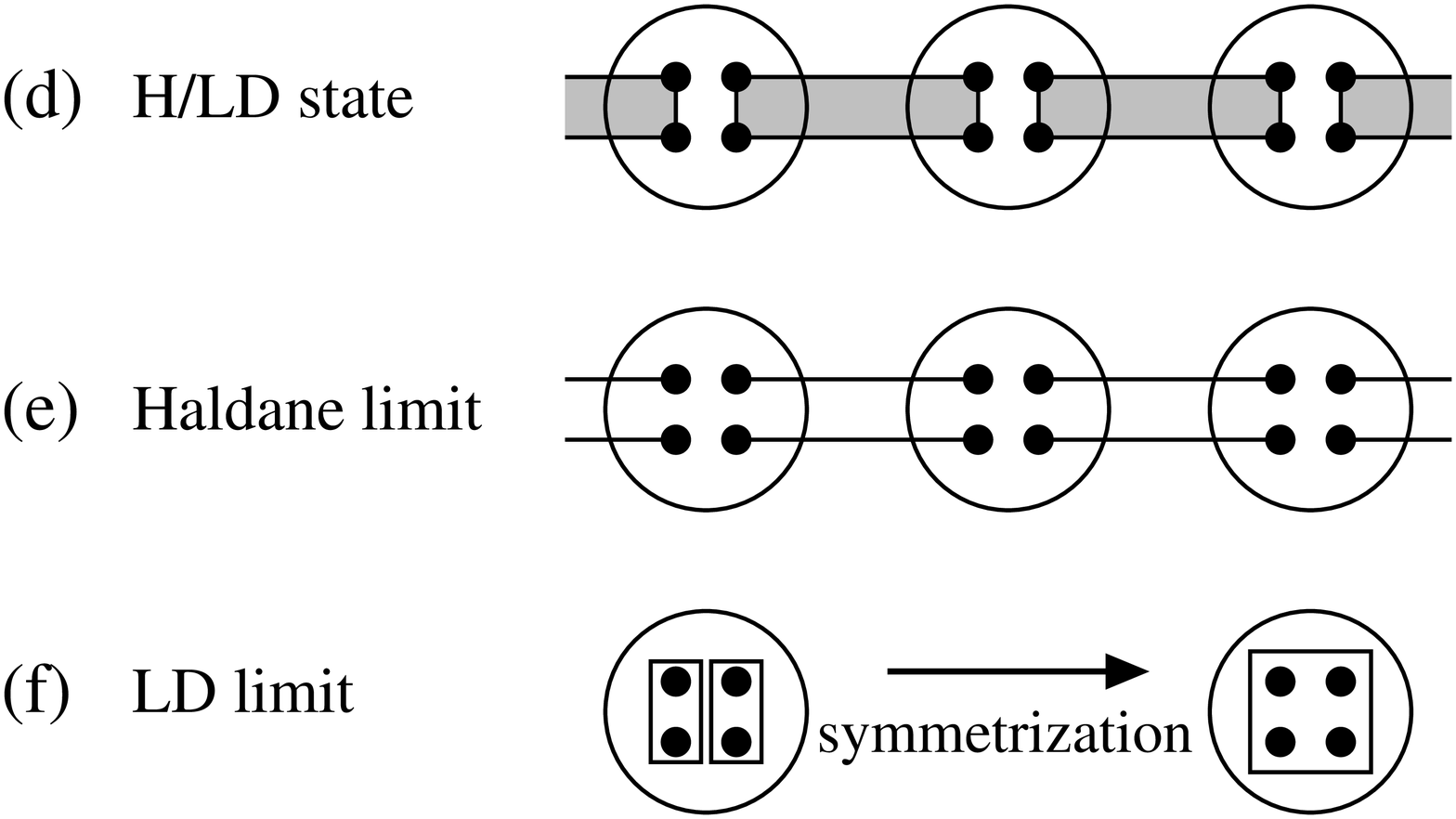}}
      \end{center}
   \caption{VBS pictures for (a) the Haldane state, (b) the ID state and (c) the LD state.
   Big circles denote $S=2$ spins and dots $S=1/2$ spins.
   Solid lines represent valence bonds (singlet pairs of two $S=1/2$ spins, 
   $(1/\sqrt{2})(\uparrow\downarrow-\downarrow\uparrow)$).
   Two $S=1/2$ spins in rectangles are in $(S_{\rm tot},S^z_{\rm tot})=(1,0)$ state
   and similarly four $S=1/2$ spins in squares are in $(S_{\rm tot},S^z_{\rm tot})=(2,0)$ state.
   We show a VBS picture of the H/LD state in (d), where the shaded rectangles denote four-spin clusters.
   The Haldane state (e) and the LD state (f) are interpreted
   as the limiting cases of the four-spin cluster state.}
   \label{fig:vbs-pictures}
\end{figure}

We have obtained the phase diagram Fig.1 on the $\Delta-D_2$ plane
by use of the exact diagonalization (ED) and the level spectroscopy (LS)
analysis \cite{ls1,ls2,ls3,ls4} as well as the phenomenological renormalization group
(PRG) analysis \cite{nightingale}.
For simplicity we have restricted ourselves to the case where $\Delta \ge 0$ and $D_2 \ge 0$.
In this phase diagram there are four phases:
the XY1 phase, the intermediate-$D$ (ID) phase, Haldane/large-$D$ (H/LD) phase 
and the N\'eel phase.
The XY1 state is gapless and characterized 
by the power decay of the spin correlation function
$G_{\perp 1}(r) \equiv \ave{S_1^+ S_r^-}$.
The Haldane, ID and LD states are gapful and their valence bond solid (VBS) pictures
are shown in Fig.\ref{fig:vbs-pictures}(a)-(c).
The remarkable natures of this phase diagram are as follows:
(A) there exists the ID phase which was predicted by Oshikawa about twenty years
ago \cite{oshikawa};
(B) the Haldane state and the LD state belong to the same phase
which is called the H/LD phase.
At a glance,
the Haldane state and the LD state differ from each other.
The reason why these two states are essentially the same is explained in Fig.\ref{fig:vbs-pictures}(d)-(f).
The H/LD state is the four-spin cluster state,
both limits of which are interpreted as the Haldane state and the LD state, respectively.
Very similar situations are found in the $S=1$ chain
with the on-site anisotropy and the bond alternation \cite{tone2},
and in the $S=1$ two-leg ladder \cite{todo}.

Slightly after our works \cite{tone1,oka1,oka2},
Tzeng \cite{tzeng} confirmed our results by use of the
density matrix renormalization group (DMRG) and the LS analysis.
He used the parity DMRG because the classification of the eigenstates by the parity
is essential for the LS analysis of the numerical data.
Kj\"all et al. \cite{kjall} also investigated this model
by use of the DMRG based on the matrix-product states
to obtain the same conclusion as ours for (B)
and slightly different conclusion from ours for (A).
Namely, they claimed that the ID phase does not exist on the $\Delta-D_2$ plane
and very small $D_4$ term (see eq.(\ref{eq:ham2}) and later) with $D_4 >0$  is necessary to realize the ID phase,
although they could not completely rule out the possibility of the
existence of the ID phase when $D_4=0$.
We think that the difference between our and their conclusions for (A) is not
a serious problem.
Since the edge of the ID phase in the $\Delta-D_2-D_4$ space near $D_4 = 0$
may be very sharp,
it is exceedingly difficult to numerically determine whether this sharp edge 
reaches the $D_4=0$ plane or not.

In the $S=1$ chain problem,
the edge modes under the open boundary condition (OBC) are very important to
elucidate and characterize the nature of the Haldane state.
In this paper, we investigate the edge modes of the H/LD state and the ID state.
However, 
since the region of the ID phase on the $\Delta-D_2$ plane, if exists, is very narrow
as can be seen from Fig.\ref{fig:pd1},  
it may be extremely difficult to obtain clear results on the edge modes.
Thus, we modify the Hamiltonian into
\begin{equation}
    \cH
    = \sum_j (S_j^x S_{j+1}^x + S_j^y S_{j+1}^y +  \Delta S_j^z S_{j+1}^z)
      + D_2 \sum_j (S_j^z)^2 + D_4 \sum_j (S_j^z)^4,
    \label{eq:ham2}
\end{equation}
by introducing the $D_4$ term to expand the region of the ID phase.


\section{Phase Diagram}

\begin{figure}[ht]
   \begin{center}
       \scalebox{0.25}{\includegraphics{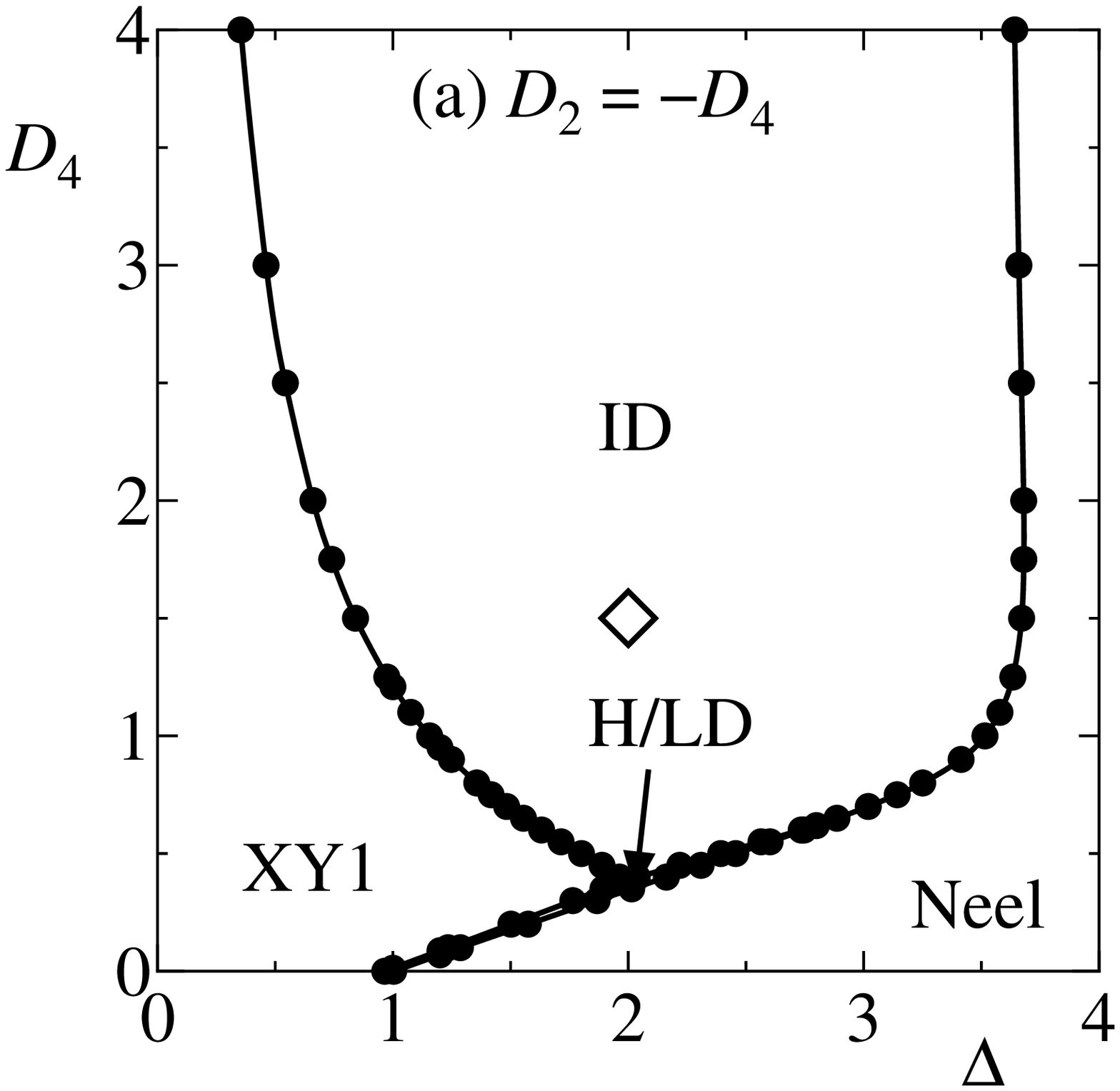}}~~~
       \scalebox{0.25}{\includegraphics{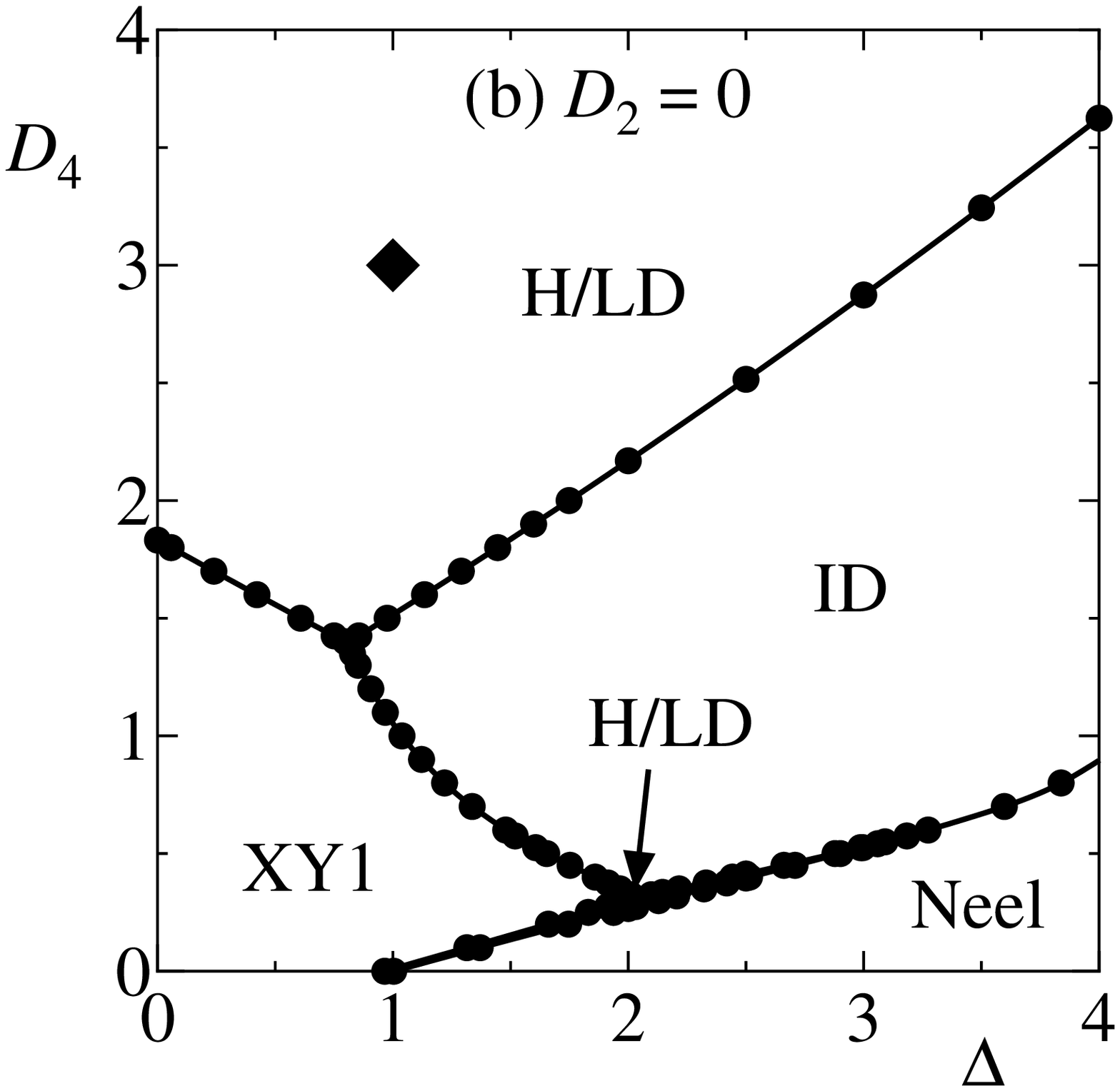}}~~~
       \scalebox{0.25}{\includegraphics{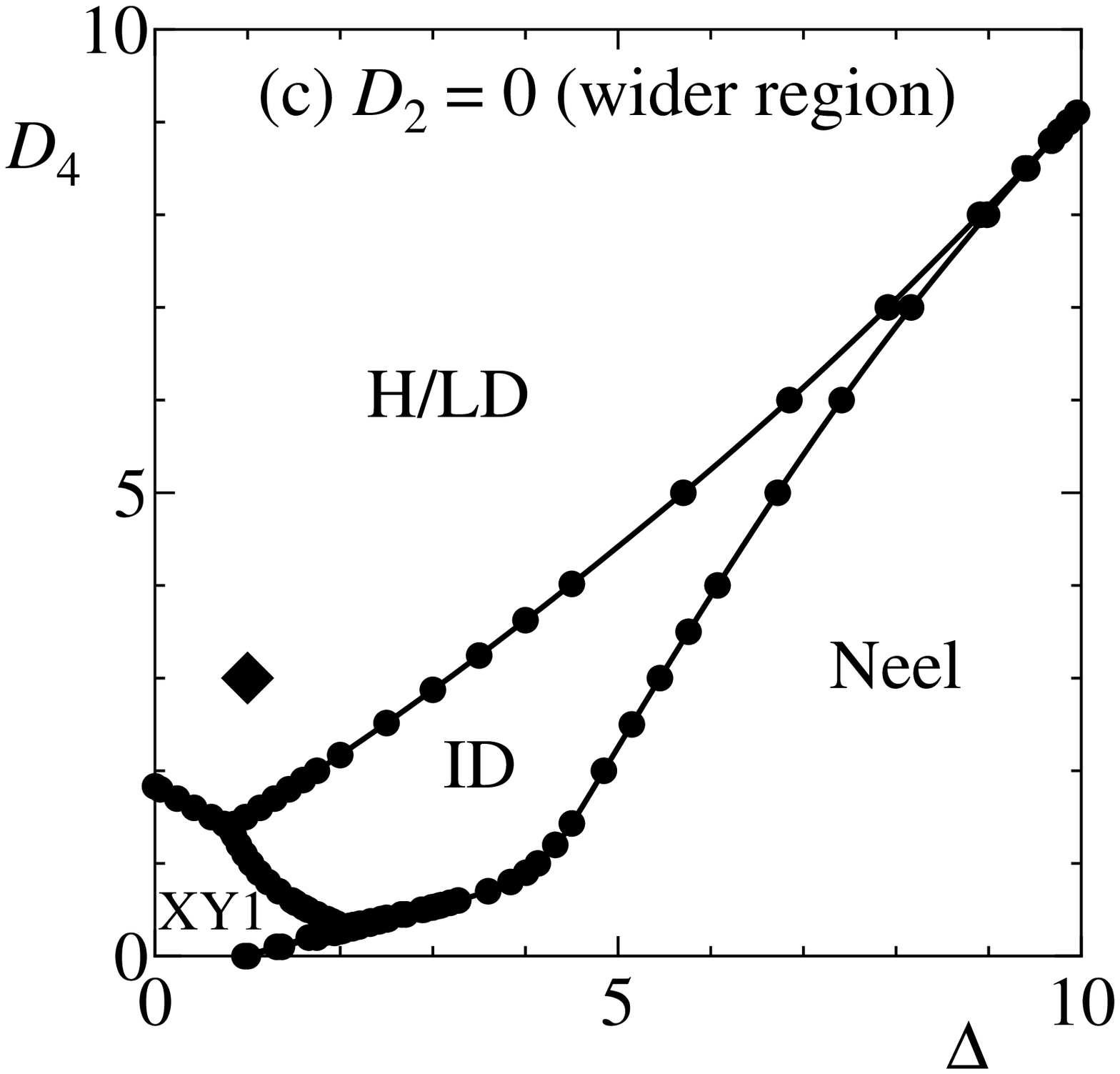}}
   \end{center}
   \caption{Phase diagram of Hamiltonian (\ref{eq:ham2}) on the $\Delta-D_4$
            plane for
            (a) $D_2 = -D_4$ and (b), (c) $D_2 = 0$. 
            For simplicity, we restrict ourselves to the case of $\Delta \ge 0$
            and $D_4 \ge 0$.
            The open and closed diamond marks (\rotatebox{45}{\footnotesize$\square$} and
            \rotatebox{45}{\footnotesize$\blacksquare$} )
            show the points where the edge mode behaviors are calculated.
           } 
   \label{fig:pd2}
\end{figure}

We mainly investigate two cases;  the $D_2 =-D_4$ case and the $D_2=0$ case.
The phase diagrams are shown in Fig.\ref{fig:pd2},
which were determined by the ED and the LS analysis as well as
the PRG analysis (see \cite{tone1,oka1,oka2} for details of the analysis).
In both cases,
the ID regions are considerably enlarged
and the narrow H/LD regions exist near the $(\Delta,D_4) =(1,0)$ point
(isotropic point).
The H/LD region at the larger $D_4$ case is completely separated
from that near the $(\Delta,D_4) =(1,0)$ point in (b) and (c),
while it does not exist for larger $D_4$ case in (a).
The ID region in (a) survives even in the $D_4 \to \infty$ limit,
the reason of which will be discussed in \S4.
The H/LD-ID transition line and the ID-N\'eel transition line
in Figs.\ref{fig:pd2}(b) and (c)
merge at the multicritical point $(\Delta, D_4) \simeq (9.5,8.5)$.
In the upper right region of this multicritical point,
there exists the H/LD-N\'eel transition line.

Kj\"all et al. \cite{kjall} also studied the model (\ref{eq:ham2}) and draw the phase diagrams.
Since they draw the phase diagrams on the $D_2-D_4$ plane for the $\Delta=1.0$ and
the $\Delta = 4.5$ cases,
we cannot directly compare our phase diagrams with theirs.
However, we can compare ours and theirs along some lines, for instance,
the $D_4$ line with $\Delta = 1.0$ and $D_2=0.0$.
On these lines, our results and theirs are consistent with each other.

\section{Edge Modes}

To investigate the behaviors of the edge modes,
we calculated the low-lying excited states by the DMRG under the OBC up to 480 sites
at the \rotatebox{45}{\footnotesize$\square$} point $(\Delta,D_2,D_4) = (2,-1.5,1.5)$
and the \rotatebox{45}{\footnotesize$\blacksquare$} point $(\Delta,D_2,D_4) = (1,0,3)$
in Fig.\ref{fig:pd2},
as representatives of the ID state and the H/LD state, respectively.
We calculated finite-size gaps with the $N$-spin system defined by
\begin{equation}
    \Delta E_{01} \equiv E_0(M=1) - E_0(M=0),~~~
    \Delta E_{02} \equiv E_0(M=2) - E_0(M=0).
\end{equation}
Here $E_0(M=0)$, $E_0(M=1)$ and $E_0(M=2)$ are the lowest energies within the subspaces 
$M=0$, $M=1$ and $M=2$, respectively,
where $M \equiv \sum_j S_j^z$ is the total  magnetization.
We also calculated the local magnetization $m(j)$ and their sum
defined by
\begin{equation}
     m(j) = \ave{S_j^z},~~~~~Q(j) = \sum_{i=1}^j m(i),
\end{equation} 
respectively.
\begin{figure}[ht]
   \begin{center}
       \scalebox{0.21}{\includegraphics{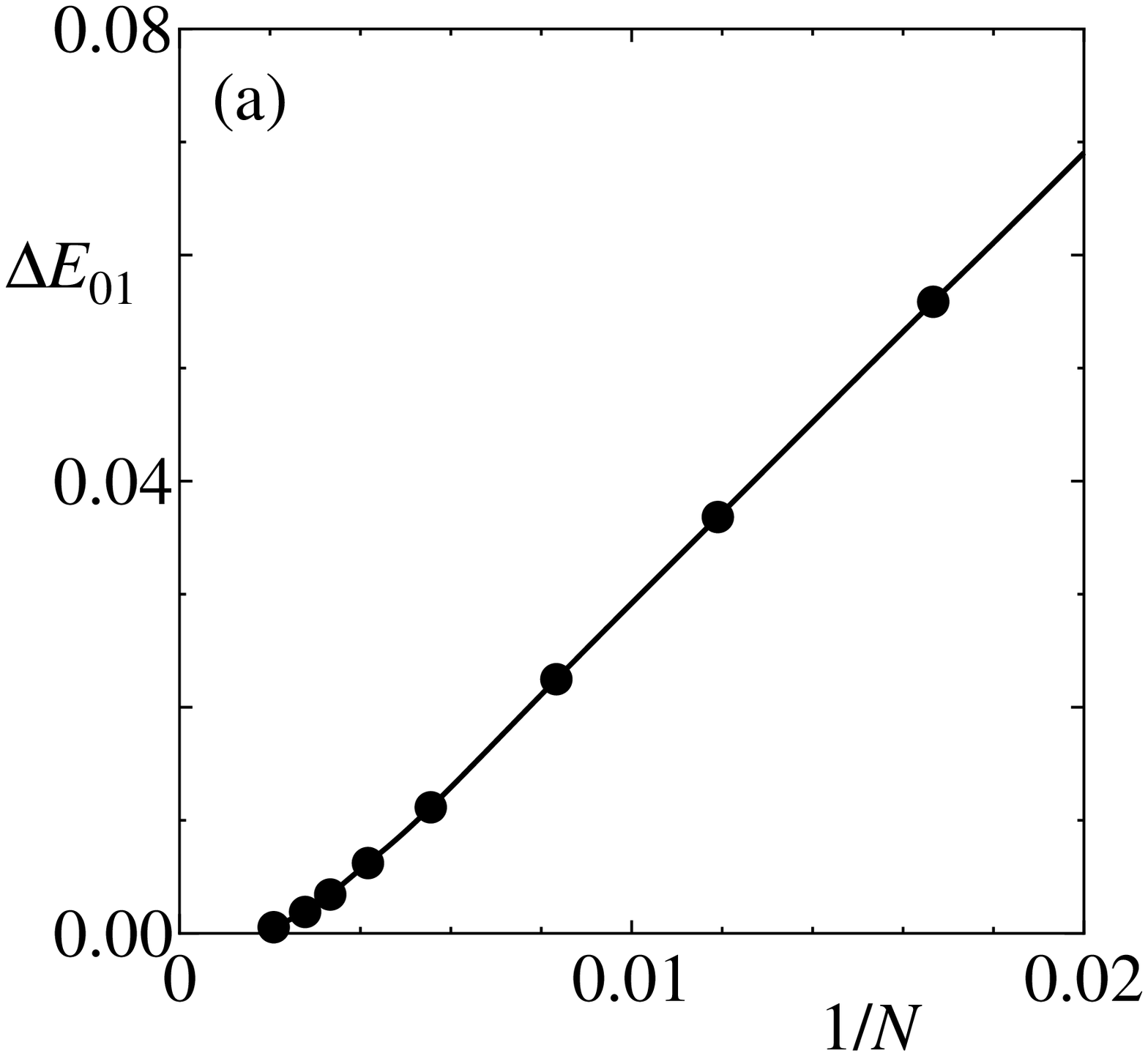}}~~~
       \scalebox{0.21}{\includegraphics{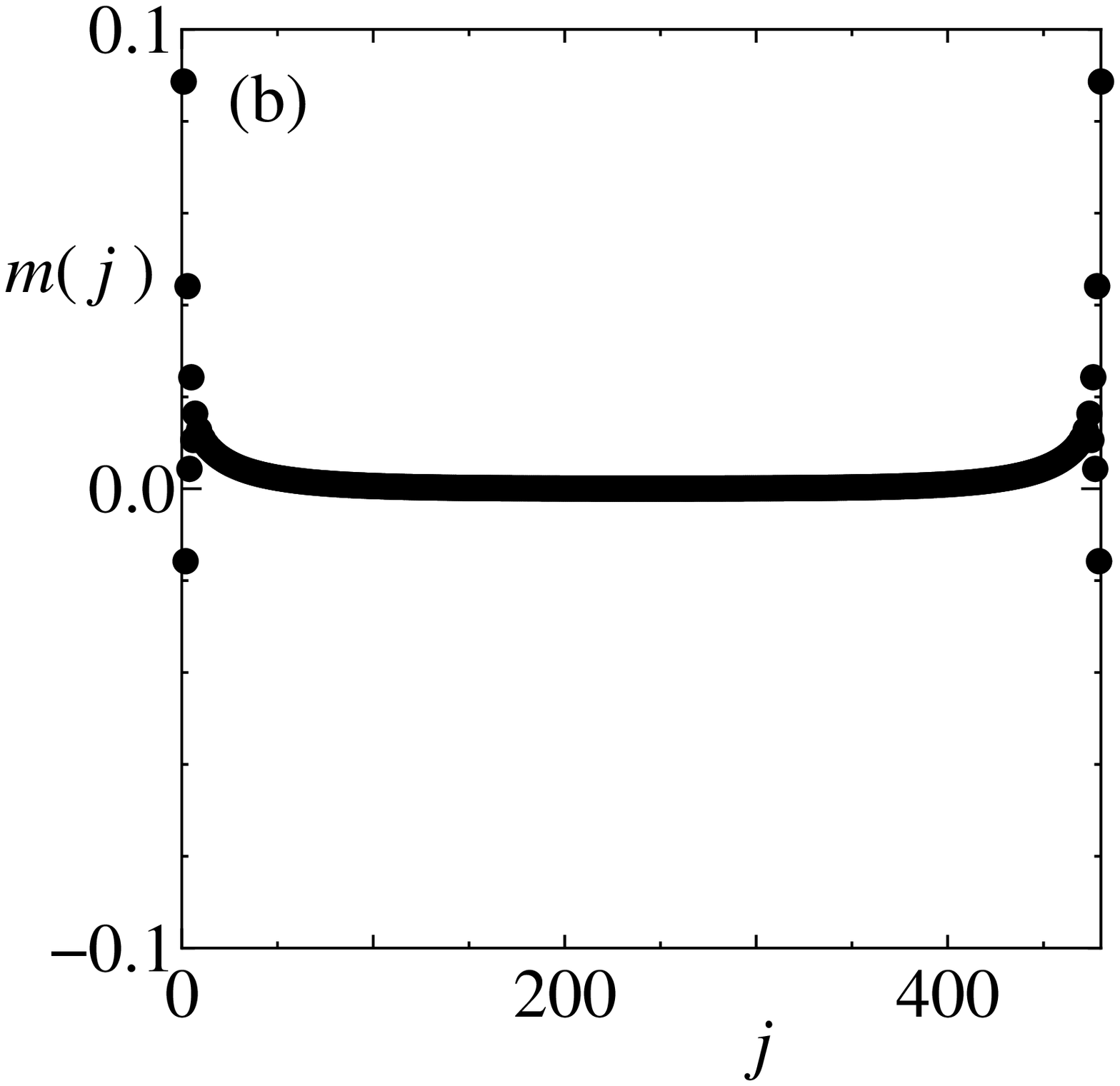}}~~~
       \scalebox{0.21}{\includegraphics{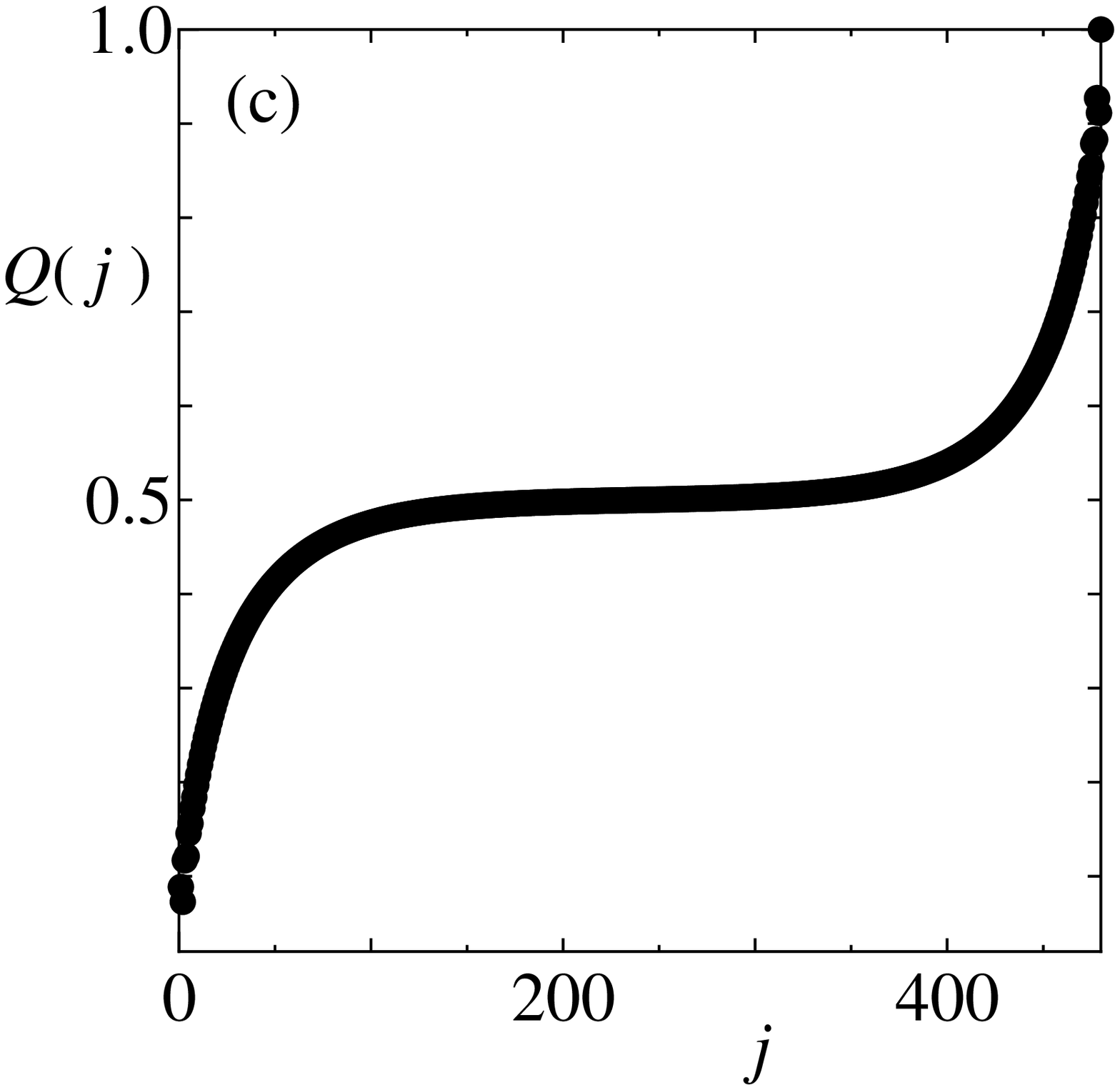}}~~~
   \end{center}
   \caption{The lowest $M=1$ mode at the \rotatebox{45}{\footnotesize$\square$}
            point (ID phase)  $(\Delta,D_2,D_4) = (2,-1.5,1.5)$.
            } 
   \label{fig:edge-ID-m1}
\end{figure}
\vskip-1cm
\begin{figure}[ht]
   \begin{center}
       \scalebox{0.21}{\includegraphics{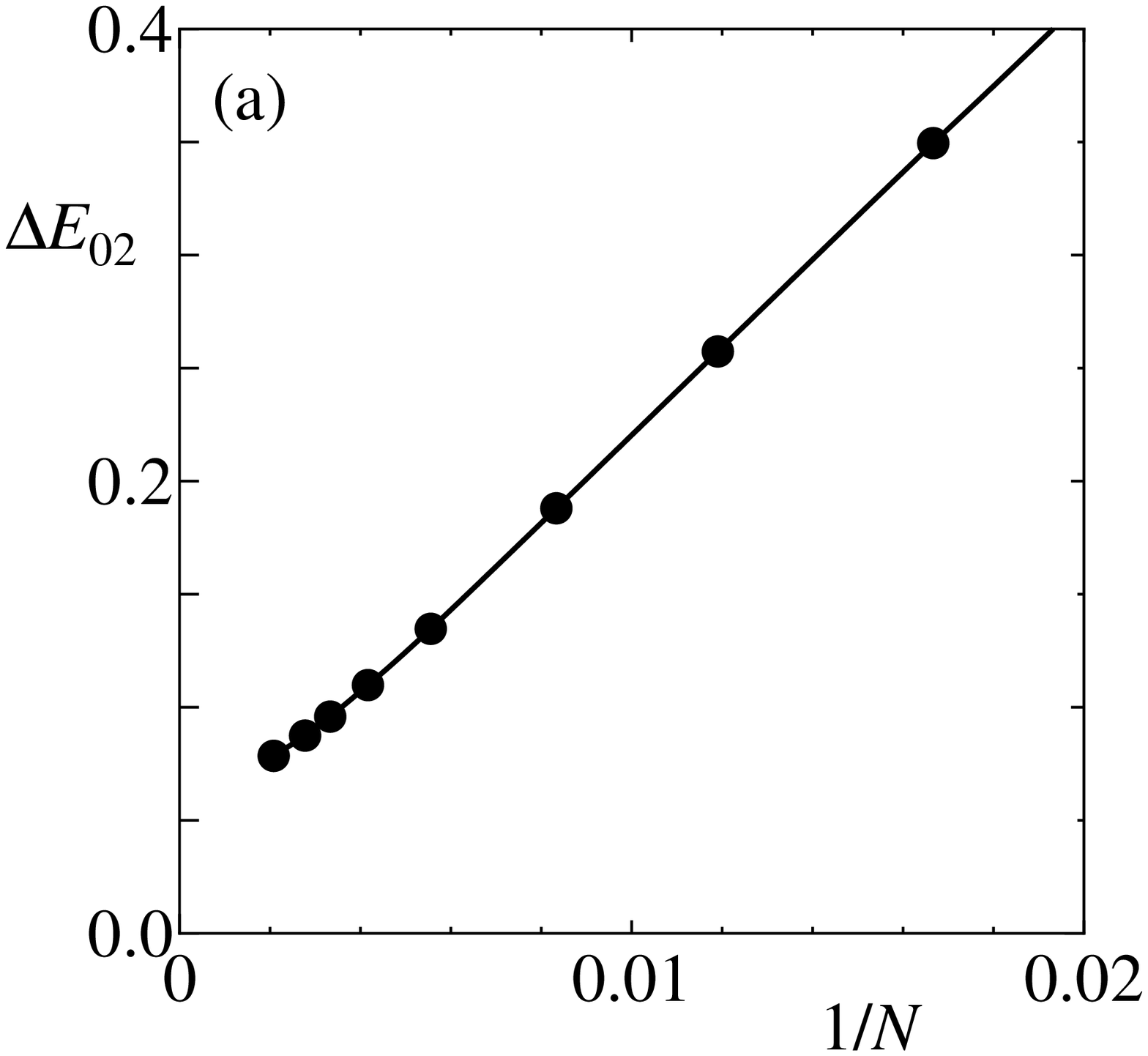}}~~~
       \scalebox{0.21}{\includegraphics{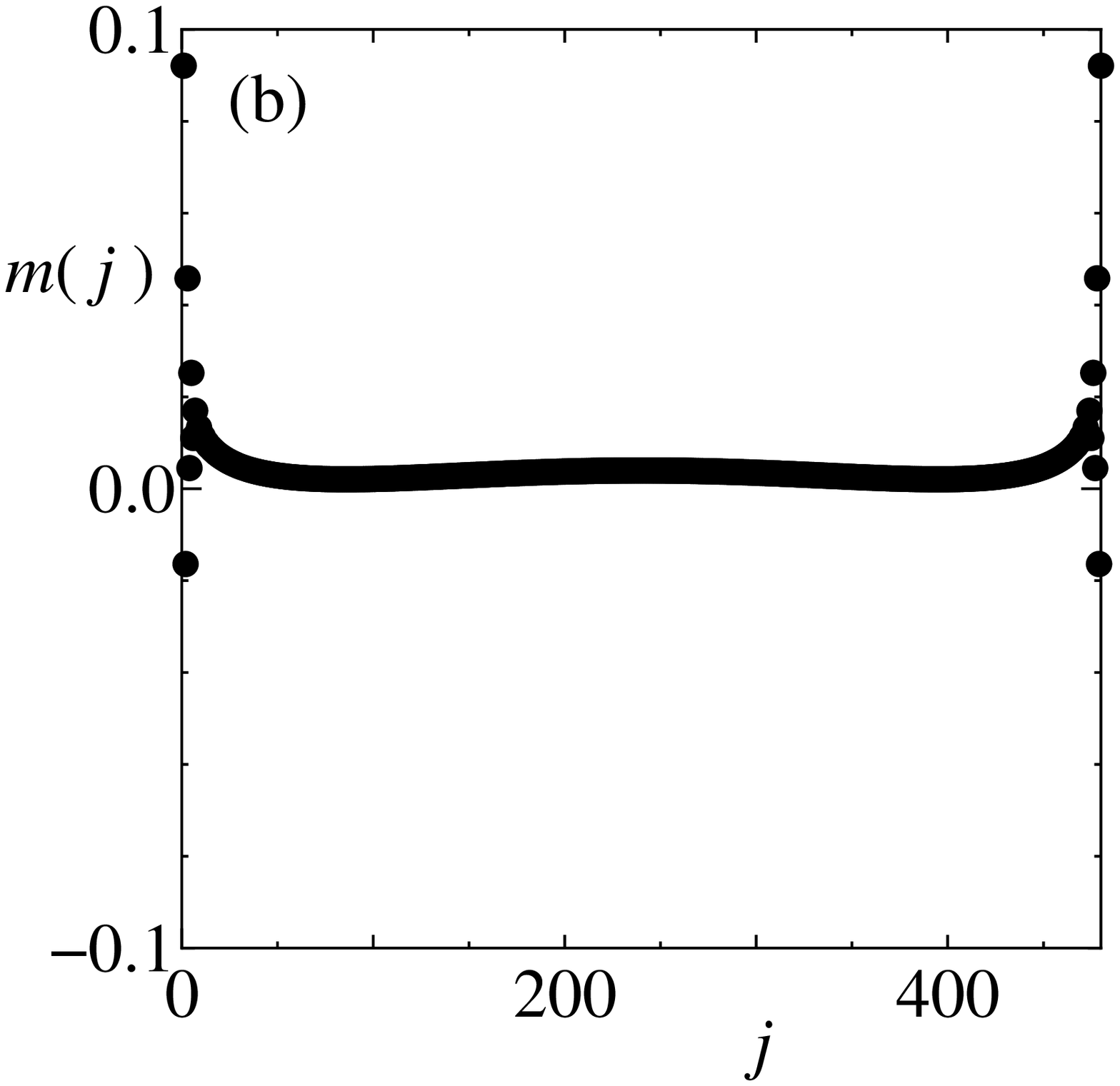}}~~~
       \scalebox{0.21}{\includegraphics{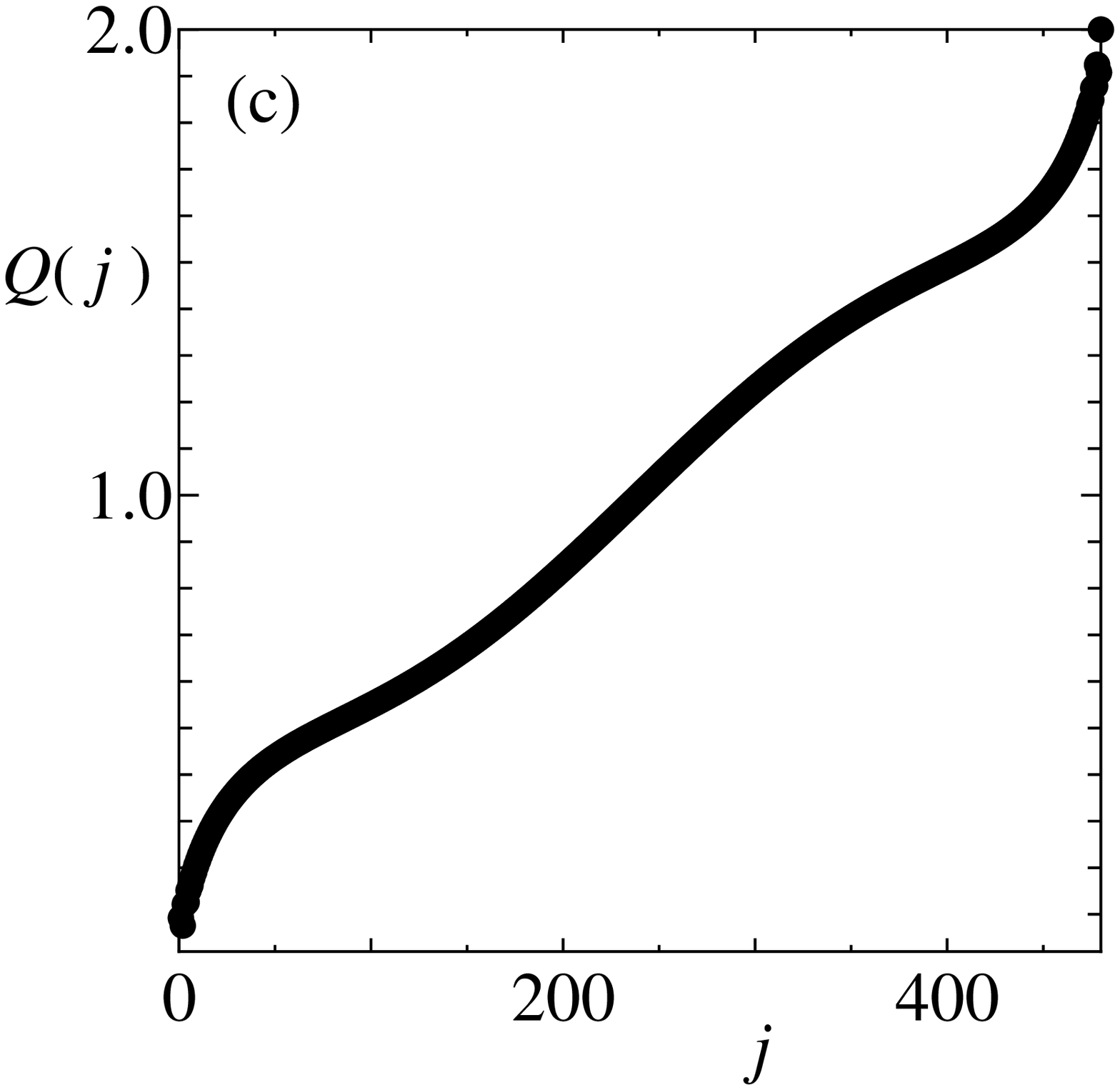}}
   \end{center}
   \caption{The lowest $M=2$ mode at the \rotatebox{45}{\footnotesize$\square$}
            point (ID phase)  $(\Delta,D_2,D_4) = (2,-1.5,1.5)$.
            } 
   \label{fig:edge-ID-m2}
\end{figure}
\vskip-1cm
\begin{figure}[ht]
   \begin{center}
       \scalebox{0.21}{\includegraphics{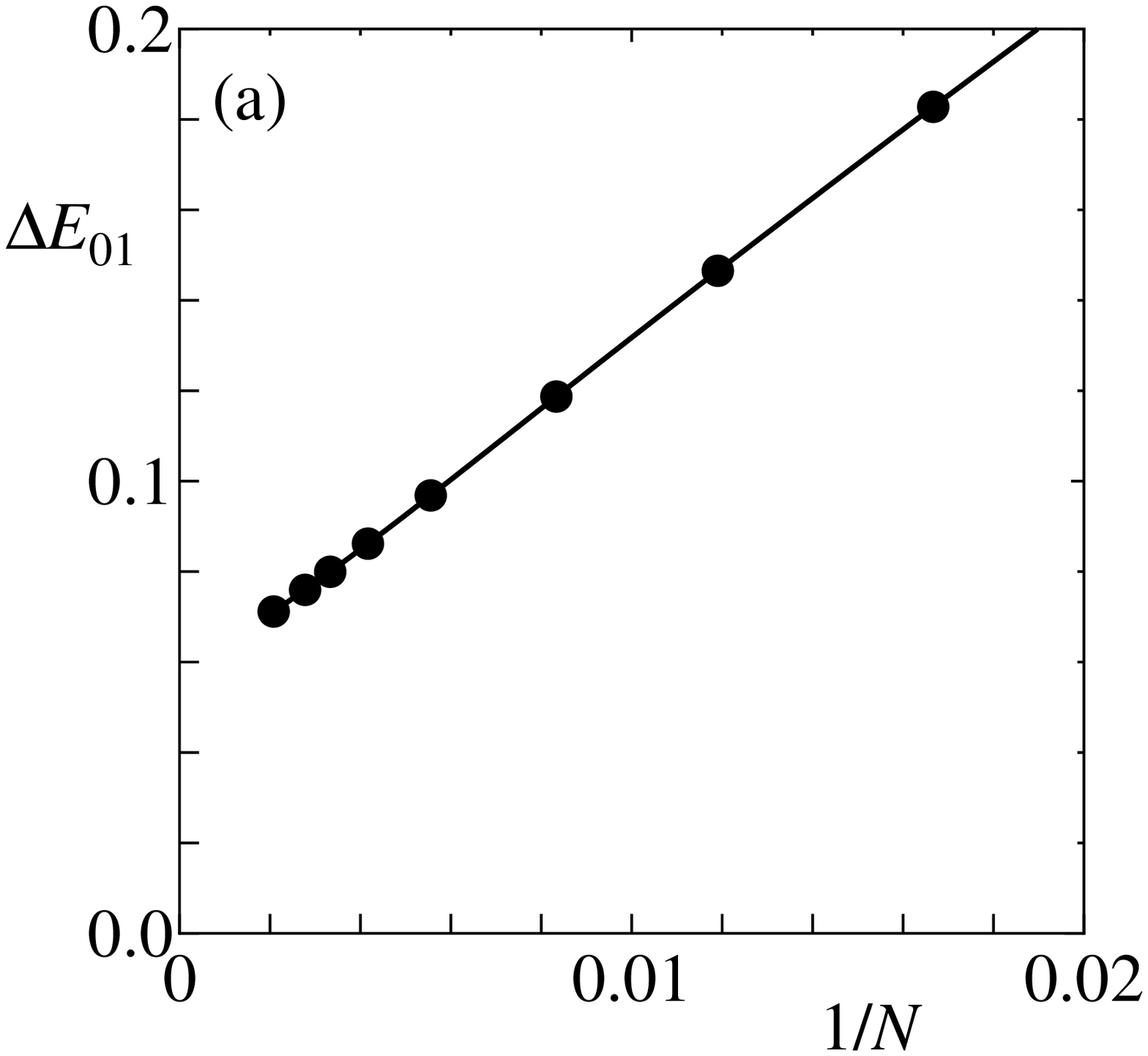}}~~~
       \scalebox{0.21}{\includegraphics{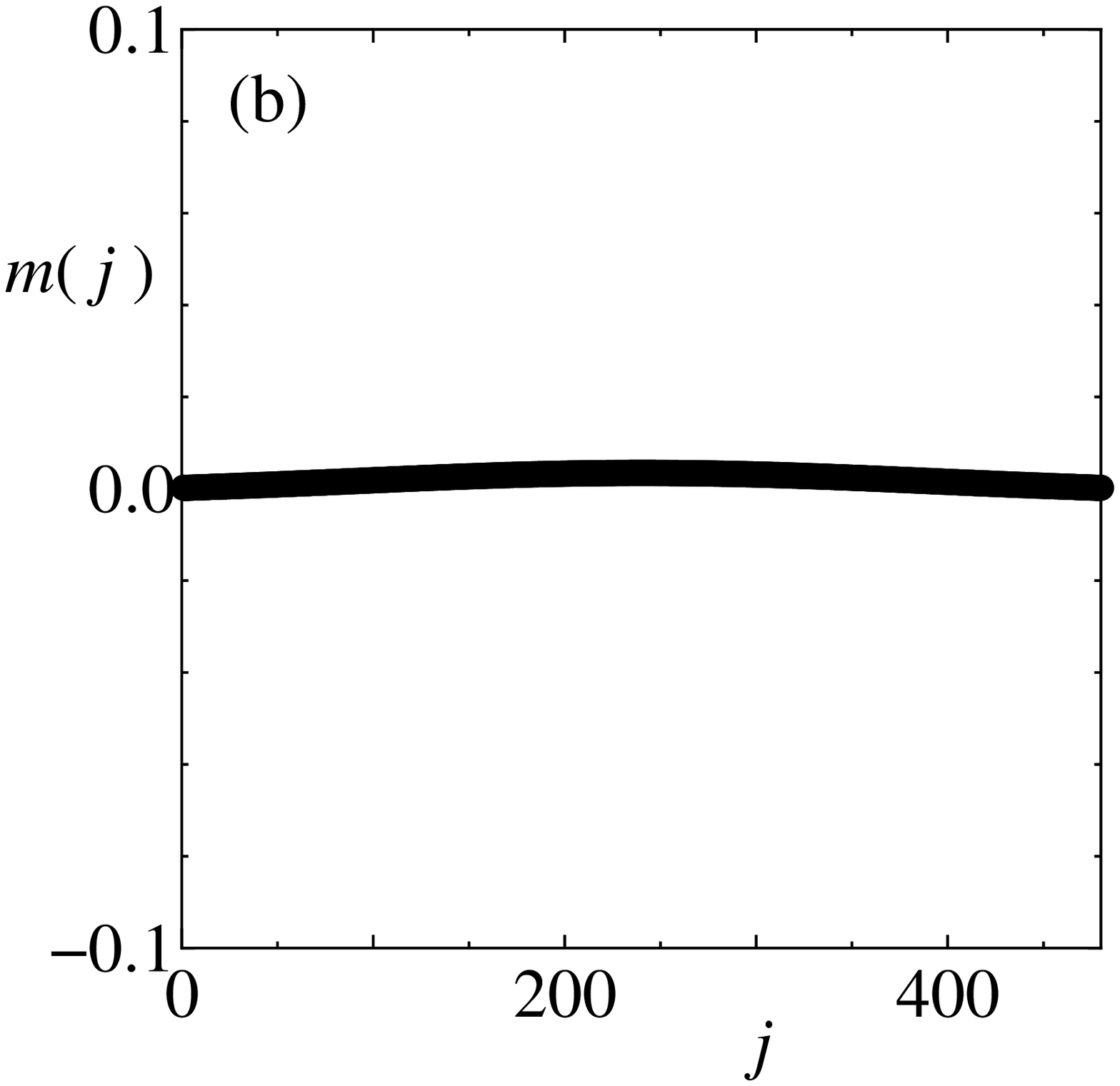}}~~~
       \scalebox{0.21}{\includegraphics{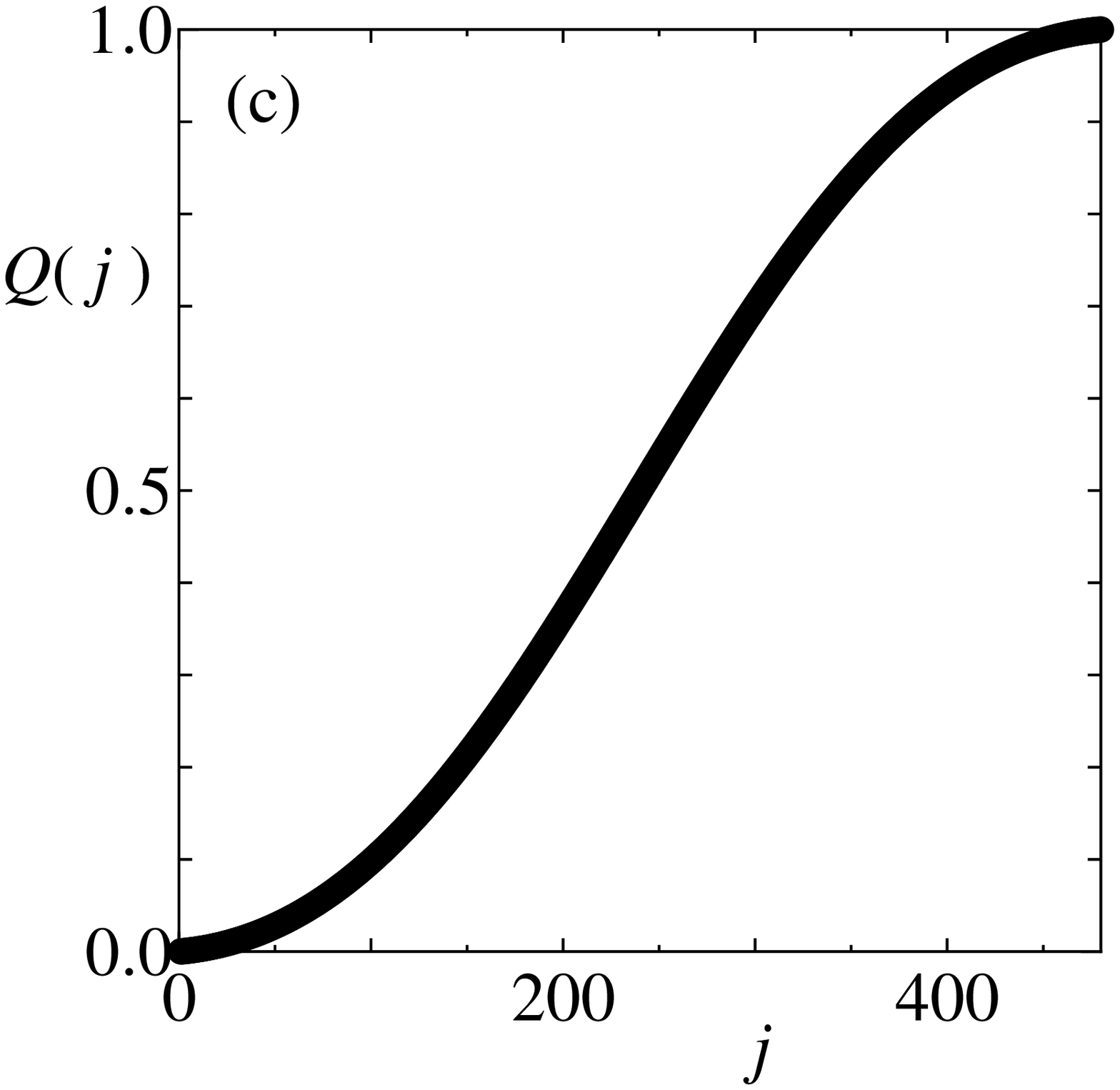}}
   \end{center}
   \caption{The lowest $M=1$ mode at the \rotatebox{45}{\footnotesize$\blacksquare$}
            point (ID state)  $(\Delta,D_2,D_4) = (1,0,3)$.
            } 
   \label{fig:edge-LD-m1}
\end{figure}

We show the results at the \rotatebox{45}{\footnotesize$\square$} point (ID state)
in Fig.\ref{fig:edge-ID-m1} (the lowest $M=1$ mode)
and in Fig.\ref{fig:edge-ID-m2} (the lowest $M=2$ mode).
We can see that the $M=1$ mode is gapless and an edge-localized mode,
while the $M=2$ mode is gapful and a bulk mode. 
Figure \ref{fig:edge-LD-m1} shows the $M=1$ mode
at the \rotatebox{45}{\footnotesize$\blacksquare$} point (H/LD state),
from which we know that this mode is gapful and a bulk mode. 
These behavior are consistent with the VBS pictures as will be discussed in \S4.

\section{Discussion}

In Fig.\ref{fig:pd2}(a),
the ID phase survives even in the $D_4 \to \infty$ limit.
In  the $D_2 = -D_4$ case,
the on-site anisotropy terms can be expressed as $D_4 \sum_j (S_j^z)^2 [(S_j^z)^2 -1 ]$.
Thus, the $S_j^z =0$ and $S_j^z =\pm 1$ states are not affected by the on-site anisotropy terms,
while the $S_j^z = \pm 2$ state is strongly suppressed for larger $D_4$.
Thus,
in the $D_4 \to \infty$ case,
we can map the present model onto the $T=1$ model by neglecting the $S_j^z = \pm 2$ states,
which results in
\begin{equation}
    \cH_\eff^{(1)}
    = 3 \sum \left( T_j^x T_{j+1}^x + T_j^y T_{j+1}^y + \Delta_\eff T_j^z T_{j+1}^z \right),~~~~~
    \Delta_\eff = {1 \over 3}\Delta.
\end{equation}
Here the Haldane state of $\cH_\eff^{(1)}$ corresponds to 
the ID state of the present $S=2$ model.
The $XY$-Haldane transition point of $\cH_\eff^{(1)}$ is exactly $\Delta_\eff = 0$
\cite{alcaraz,kitazawa1,kitazawa2,kitazawa3,kitazawa4},
and the Haldane-N\'eel transition point is $\Delta_\eff \simeq 1.23$ \cite{sakai,chen}.
These points correspond to $\Delta =0$ and $\Delta \simeq 3.69$ in our $S=2$ chain, respectively,
which well explains the behaviors of the $XY1$-ID line and the ID-N\'eel line
in the $D_4 \to \infty$ limit of Fig.\ref{fig:pd2}(a).

Also in the $D_2=0$ case,
the $S_j^z = \pm 2$ is strongly suppressed for large $D_4$.
If we project out the $S_j^z = \pm 2$ states,
we obtain
\begin{equation}
    \cH_\eff^{(2)}
    = 3 \left\{ \sum \left( T_j^x T_{j+1}^x + T_j^y T_{j+1}^y + \Delta_\eff T_j^z T_{j+1}^z \right)
      + D_{2,\eff} \sum_j (S_j^z)^2 \right\},~~~~~
    D_{2,\eff} = {1 \over 3}D_4.
\end{equation}
In this case,
the Haldane state and the LD state of $\cH_\eff^{(2)}$ correspond to
the ID state and the LD state of the present $S=2$ model, respectively.
The multicritical point among the Haldane, LD and N\'eel phases of $\cH_\eff^{(2)}$ 
is $(\Delta_\eff,D_{2,\eff}) \simeq (3.2,2.9)$ \cite{chen},
which reads $(\Delta,D_4) \simeq (9.4,8.7)$.
Thus the multicritical point among the ID, LD and N\'eel phases
in Fig.\ref{fig:pd2}(c) is well explained.

\begin{wrapfigure}[10]{r}{9cm}
   \scalebox{0.3}{\includegraphics{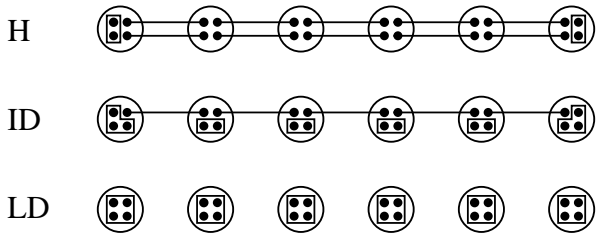}}
   \caption{VBS pictures with edges for the Haldane (H) state, the ID state and the LD state.
   }
   \label{fig:vbs-pictures-edge}
\end{wrapfigure}
Figure \ref{fig:vbs-pictures-edge} shows the VBS pictures
with edges for the Haldane (H) state, the ID state and the LD state.
For the Haldane state, there appear $S=1$ edge spins.
In the presence of the on-site anisotropy,
the $S^z = \pm 1$ edge states have higher energies than the $S_z=0$ state.
The lowest $M=0$ mode can be made by use of two $S^z =0$ edge spins.
On the other hand,
the lowest $M=1$ edge mode consists of the $S^z = 0$ edge spin and the $S^z = 1$ edge spin,
which means that the $M=1$ edge mode is gapful.
We note that, in the case of $D_2=D_4=0$,
we can construct the gapless edge modes with $M=1$ and $M=2$
without the loss of on-site energy,
because $S_z = 0$ and $S^z = 1$ edge spins have same energies,
as was numerically obtained by Qin et al. \cite{qin}.

For the ID state, there appear $S=3/2$ edge spins.
The $S^z = \pm 1/2$ states have lower energies than the $S^z= \pm 3/2$ states.
The lowest $M=0$ mode is constructed by use of an $S^z = +1/2$ edge spin and
an $S^z =-1/2$ edge spin,
while the lowest $M=1$ mode by use of two $S^z = +1/2$ edge spins.
Thus, there appears the $M=1$ gapless edge mode in the ID phase.
Because we have to use an $S^z = 3/2$ edge state for constructing an $M=2$ edge mode,
there is no $M=2$ gapless edge mode in the ID phase.
Thus the calculated behaviors in Figs.\ref{fig:edge-ID-m1} and \ref{fig:edge-ID-m2}
are successfully explained by use of the VBS pictures.

About the LD state, to construct an edge $M=1$ mode,
it is necessary to change the $S^z=0$ state of either edge spin to the $S^z=1$ state,
which brings about the energy loss of $D_4$ in the case of $D_2=0$, $D_4>0$.
Thus it is clear that the $M=1$ mode is gapful,
which is consistent with Fig.\ref{fig:edge-LD-m1}.

In conclusion,
we have investigated the properties of edge modes in the ID and LD states
of the $S=2$ quantum spin chain with the $XXZ$ and the generalized on-site anisotropies.
These are well explained by use of the VBS pictures.

\section*{Acknowledgements}

We would like to express our appreciation to 
Masaki Oshikawa for his invaluable discussions and comments as well as for his
interest in this study. 
We are deeply grateful to Frank Pollmann for his
stimulating discussions. We also thank the Supercomputer Center, Institute for
Solid State Physics, the University of Tokyo, and the Computer Room, Yukawa Institute
for Theoretical Physics, Kyoto University.
This work was partly supported by Grants-in-Aid (Nos. 23340109 and 23540388)
for Scientific Research from the Ministry of Education, Culture, Sports, Science and Technology of Japan.

\end{document}